\documentclass[
aps,
floatfix,
graphix, 
amsmath,
amssymb,
reprint]{revtex4-1}
\usepackage{graphicx}
\usepackage{dcolumn}
\usepackage{bm}
\usepackage{amsmath} 
\usepackage{bbold} 
\usepackage{appendix}

\begin{document}

\title{Multitask machine learning of collective variables for enhanced sampling of rare events}

\author{Lixin Sun}
\author{Jonathan Vandermause}
\author{Simon Batzner}
\author{Yu Xie}
\author{David Clark}
\author{Wei Chen}
\author{Boris Kozinsky}
\email{bkoz@seas.harvard.edu}
\affiliation{
John A. Paulson School of Engineering and Applied Sciences, Harvard University
}

\begin{abstract}
Computing accurate reaction rates is a central challenge in computational chemistry and biology
because of the high cost of free energy estimation with unbiased molecular dynamics.
In this work, a data-driven machine learning algorithm is devised to learn collective variables with a multitask neural network, where a common upstream part reduces the high dimensionality of atomic configurations to a low dimensional latent space, and separate downstream parts map the latent space to predictions of basin class labels and potential energies.
The resulting latent space is shown to be an effective low-dimensional representation, capturing the reaction progress and guiding effective umbrella sampling to obtain accurate free energy landscapes.
This approach is successfully applied to model systems including a 5D M{\"u}ller Brown model, a 5D three-well model and alanine dipeptide in vacuum.
This approach enables automated dimensionality reduction for energy controlled reactions in complex systems, offers a unified framework that can be trained with limited data, and outperforms single-task learning approaches, including autoencoders.
\end{abstract}


\maketitle

\section{Introduction}

Computing accurate reaction rates is one of the most important challenges
in computational physics, chemistry and biology. 
Reactions are rare events in which a system transitions from one metastable state to another.
Reactions with high barriers can have time scales of microseconds or longer, 
such that conventional unbiased molecular dynamics (MD) is too slow to accumulate enough statistics on transitions to calculate accurate reaction rates.
Enhanced sampling methods, such as umbrella sampling\cite{torrie1977nonphysical, souaille2001extension} and metadynamics\cite{barducci_metadynamics_2011, bonomi_reconstructing_2009}, 
address this challenge by accelerating sampling of phase space using biasing applied along low-dimensional collective variables (CVs) direction. 
These methods require one to first reduce the high-dimensional configuration space to a low-dimensional manifold of CVs in order to evaluate the free energy landscape
where metastable states correspond to local minima basins and transition states correspond to
high free energy separation ridges.
The reaction rate can then be estimated within the transition state theory framework.

Good CVs need to discern different metastable states and transition paths;
they can be simple geometrical variables such as atomic coordination numbers\cite{ensing_perspective_2005}, or combinations of bond distances and bond angles\cite{baftizadeh_protein_2012}.
Reaction coordinates (RCs) are a type of CV that need to be one dimensional that strictly preserve reaction progress
from one metastable state to another metastable state.
Linear combinations of simple geometrical variables are usually not sufficient when the transition paths are complex, and poor choices of CVs result in inefficient sampling and inaccurate reaction rates.
However, designing good CVs is a laborious trial-and-error process typically requiring intuition and prior knowledge of the relevant reaction mechanisms \cite{torrie1977nonphysical, souaille2001extension, barducci_metadynamics_2011, bonomi_reconstructing_2009}.

In the spirit of data-driven analysis, a number of methods have recently been employed to design CVs and RCs with machine learning (ML) \cite{rydzewski_machine_2016, boninsegna_investigating_2015, ferguson_nonlinear_2011, ceriotti_simplifying_2011, ardevol_probing_2015, ardevol_general_2016, chen_collective_2018, wehmeyer_time-lagged_2018, tiwary_spectral_2016, branduardi_b_2007, hovan_defining_2019, jung_artificial_2019, li_computing_2019, mendels_collective_2018, sultan_automated_2018}, but reactions with high barriers are often difficult to analyze with these ML approaches. 
For example, methods that identify slow CVs separate slow motions from fast vibrations by monitoring structural evolution over time\cite{boninsegna_investigating_2015, ferguson_nonlinear_2011, ardevol_probing_2015, ardevol_general_2016}
or by clustering metastable states\cite{noe2017collective, chen2019capabilities}.
However, these methods are not practical for high-barrier reactions
because their training requires long MD trajectories with a sufficient number of transitions. These transitions are hard to obtain for high-barrier reactions unless enhanced sampling techniques with CVs are used in the first place.
This chicken-and-egg problem requires the development of iterative adaptive approaches.

There are also methods that learn one-dimensional RCs $\xi(\mathbf{x})=f(q(\mathbf{x}))$ from the committor function $q(\mathbf{x}) \in [0, 1]$ which describes the progression of a reaction between two pre-defined basins A and B\cite{dellago_transition_2003, holm_transition_2009, li_computing_2019}.
The training data typically come from transition path sampling (TPS) 
\cite{lipkowitz_trajectory-based_2010, dellago_transition_2003, bolhuis_transition_2002}
or relaxation trajectories, in which most configurations are near the transition state and have relatively high free energies.
These methods maximize a likelihood of the committor function in the transition path ensemble\cite{jung_artificial_2019} or transition state ensemble\cite{Best_Hummer_2005,Ma_Dinner_2005}.
Because the committor function perfectly preserves the reaction progress,
it is often considered the ideal reaction coordinate and used to grade the quality of other reaction coordinates\cite{Peters_2006, Best_Hummer_2005}.

However, most committor learning frameworks have only been applied to diffusive processes, where the committor function changes smoothly from 0 to 1. 
For high-barrier reactions, the committor function has a sharp change from 0 to 1 around the transition hypersurface,
while in the majority of the phase space, the committor is close to 0 or 1.
This makes it difficult to accurately estimate the committor function and the CVs derived from it, as discussed in more detail in Section IIC.

Essentially, these two groups of methods both suffer from limited training data or slow convergence in statistics for high-barrier reactions.
Therefore, there are two requirements for learning good CVs.
First, the method needs to learn from a limited amount of training data, including configurations from basins and transition states.
Second, the CVs need to capture the distinction and progression of intermediate states along the reaction path.

We introduce an approach to simultaneously fulfill the above goals using a multitask machine learning model. 
Multitask machine learning models consist of a common upstream part that processes the input data and separate downstream parts that produces several outputs, where a joint loss function is used for training.
The output of the common upstream part is called latent space.
The latent space of the optimized model will then encode the union of all the downstream information.
This training strategy has been used in the field of image classification\cite{zhang_aet_2019, gidaris_unsupervised_2018} and natural language processing\cite{collobert2008unified} for dimensionality reduction and improving generalization performance.

In this work, we represent the simultaneous requirements of the CVs as multiple loss functions, 
design separate downstream parts for each loss function,
and use the latent space as CVs.
Unlike previous methods, which typically ignore potential energies,
here the multitask learning exploits the potential energy label and uses it as a way to measure reaction progress for high-barrier reactions.
The model is trained with a combination of short MD trajectories, including relaxation from the transition state to the basins and ones that are confined to the basins with no transitions.
The learning algorithm is applied to several model systems, including a 5D M{\"u}ller-Brown model, 5D three-well model and alanine dipeptide.
The latent space is shown to be an effective low-dimensional representation of atomic configurations,
identifying the important dimension for the reactions and yielding accurate reaction free energies.
In addition, the multitask learning framework is shown to outperform single-task learning frameworks, including autoencoder.

\section{Multitask Learning Approach}

\subsection{Architecture}
In the multitask learning framework, 
both the network architecture and the loss function should be designed to reflect the training data's nature and the RC/CV learning objectives.
We can break CV learning into three tasks:
(T1) dimension reduction, (T2) separating basins, and (T3) preserving atomic structural evolution from basins to TS hypersurface.
The multitask neural network contains three parts corresponding to these three tasks.

An encoder is designed as the common upstream part to handle T1.
This encoder is a neural network whose hidden layers have progressively fewer nodes as the layer is closer to the latent layer output.
It takes as input the Cartesian coordinates of atomic configurations $\mathbf{x}$ and maps them to a low-dimensional latent space $\mathbf{\xi}$.
For T2, as discussed in section IIC, we assign a basin label $n$ to each $\mathbf{x}$, so that one of the downstream network is a classifier trained with supervised learning.
And for T3, the potential energy lables are exploited.
In a high-barrier reaction, the system has to go through low
potential energy states before climbing to the higher potential energy transition states.
Therefore, potential energies $V$ can be used as an indicator of the reaction progress.
Later discussion in section IV will show that potential energy outperforms geometry-based indicator.
Here, the second downstream part is a network that predicts potential energy.

\begin{figure}[h]
\caption{Illustration of the multitask neural network.
The input $\mathbf{x}$ is the Cartesian coordinates of the atomic configuration.
The latent space $\mathbf{\xi}$ is the output of the encoder.
The classifier and the PE networks predict the basin label $\tilde{\mathbf{n}}$ and potential energy $\tilde{V}$ from the latent space output.
\label{fig:illustration}}
\includegraphics[width=\linewidth]{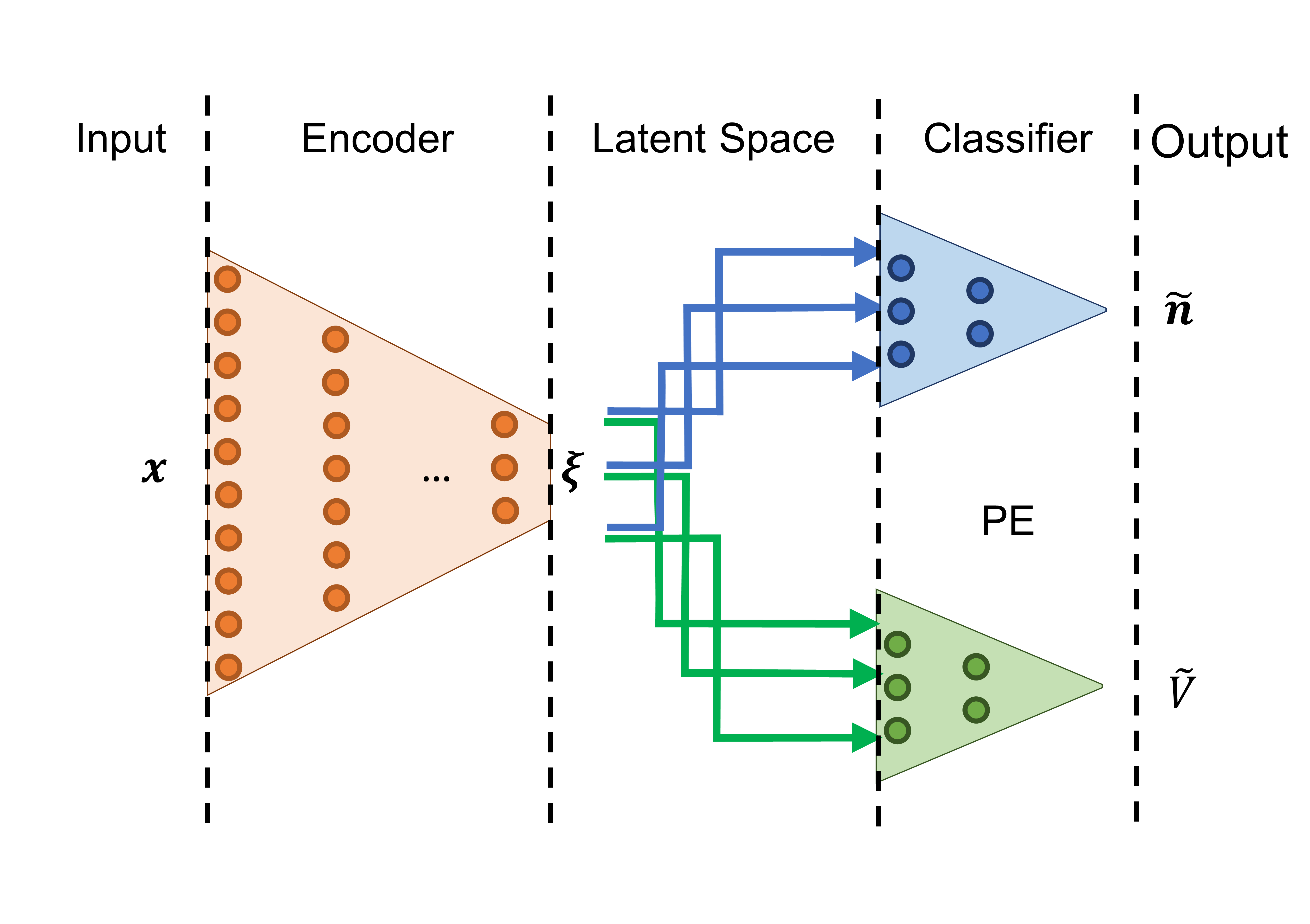}
\end{figure}

Therefore, the multitask neural network has three separate networks: an encoder, a potential energy predictor (PEP), and a basin classifier, as shown in Fig. \ref{fig:illustration}.
The encoder maps the $\mathbf{x}$ to the latent space 
$\mathbf{\xi} = f\left(\mathbf{x}\right)$.
From the latent space $\mathbf{\xi}$, the PEP predict the potential energy 
$\tilde{V}=h\left(\mathbf{\xi}\right)$ 
and the classifier network predicts the basin label encoded via a one-hot vector
${\tilde{\mathbf{n}}}=[\tilde{n_{\alpha}}]=\mathbf{g}\left(\mathbf{\xi}\right)$,
where $\alpha$ denotes the basin label and $\tilde{n}_{\alpha} \in [0, 1]$. 
In the following, the tilde sign is used to indicate predicted quantities, 
and no tilde is used to indicate actual true quantities.

\subsection{Learning objectives}
The learning objective $L$ is a joint loss function that sums several loss functions $L_p$ with coefficients $c_p$. 
\begin{align} 
\label{eq:pemse}
L =& \sum_{p} c_pL_p, \quad\quad p=\mathrm{clf}, \mathrm{pe}, \mathrm{reg} \dots; \\
\label{eq:crossentropy} L_\mathrm{clf} = & -\sum_{i} w_i \left[ n_i\log\left (\tilde{n}_i)\right) + (1-n_i)\log\left (1-\tilde{n}_i\right) \right ]; \\
\label{eq:pe_loss1} L_\mathrm{pe} = &\sum_{i} u_i (\tilde{V}_i-V_i)^2 
\end{align}
These components are the
classification error $L_\mathrm{clf}$, potential energy error $L_\mathrm{pe}$,
and regularization on the encoder weights $L_\mathrm{reg}$. 
All terms in Eq. \ref{eq:crossentropy}-\ref{eq:pe_loss1} sum over all atomic configurations $i$ in the training data.

The classification error uses the cross-entropy loss, as in Eq. \eqref{eq:crossentropy}.
This term is used to guarantee that different basins are linearly separable in the latent space.
The weights of configurations $w_i$ are normalized such that their sum is unity for each basin.

Although the above loss function is written for a two-basin scenario, 
assuming the true basin label to be integer $n=0, 1$
and the predicted label as a scalar $\tilde{n}\in [0, 1]$,
it can be extended to multiple metastable states 
with multi-class classification cross entropy.

The potential energy loss term is used to preserve the reaction progress 
from low energy basins to high energy transition states in the latent space.
Here, we choose the form for the potential energy loss as the L2 norm.
In Eq. \ref{eq:pe_loss1}, the weights $u_i$ are adjusted such that each potential energy interval
$\left[ E_\mathrm{min}, E_\mathrm{min}+\Delta E\right ]$, 
$\left[ E_\mathrm{min}+\Delta E, E_\mathrm{min}+2\Delta E\right ]$, 
$\dots$, 
$\left[ E_\mathrm{max}-\Delta E, E_\mathrm{max}\right ]$ 
contributes equally to the loss function.

\subsection{Basin classification learning}

While obtaining the potential energy labels $V$ is straightforward,
obtaining the basin labels is not trivial, because the transition state (TS) hypersurface is not known in advance.

As mentioned in the introduction, maximizing the likelihood related to the committor function can be used to learn a RC $\xi(\mathbf{x})=f(q(\mathbf{x}))$ \cite{Peters_2006, jung_artificial_2019}.
In these methods, only two-basin reactions are concerned.
For each TPS shooting point configuration $\mathbf{x}$, the probability of starting at $\mathbf{x}$ and arriving first at the $n=1$ basin is defined as the committor function 
\begin{equation}
    p_1(n=1|\mathbf{x}) \equiv q(\mathbf{x}).
    \label{eq:originalp}
\end{equation}
The committor can be estimated by maximizing the join likelihood $\mathcal{L}=\prod_i p_1(n=1|\mathbf{x}_i)$ over all training data $i$.

Mathematically, the loss function of these methods is equivalent to the cross-entropy classification loss. The committor estimator can hence be a basin classifier that classifies atomic configurations by the predicted basin labels $\tilde{n}(\mathbf{x})=q(\mathbf{x})$ (see the proof in Supplementary Section E).
By definition, the classification boundary $\tilde{n}=0.5$ is the transition state hypersurface separating the two basins.

But these frameworks are very data-inefficient for high-barrier reactions.
Only the shooting point configurations are utilized which constitute less than 1\% of all computed configurations.
And it is also hard to statistically estimate the committor function in the remaining region where $q$ is close to 0 (or 1).
It requires $\sim 1/q$ (or $\frac{1}{1-q}$) trajectories for an good estimation of $q$.
Otherwise, with only a small number of trajectories, the estimator cannot discern any slight change of $q$, because the labels for the same $\mathbf{x}$ will be either all zero or all one.

In this work, all computed configurations, regardless of shooting or non-shooting point configurations, are used.
In the presence of a high energy barrier, we can utilize the considerations that 
(1) the reaction transitions do not occur in a short MD simulation 
and (2) basin recrossing rarely happens in a short relaxation trajectory starting near the TS.

Therefore, in an unbiased MD simulation, trajectories are trapped at the basin, and all configurations are labeled by the corresponding starting basin.
For short relaxation trajectories, each configurations is labeled by the ending basin.
For example, a one-way shooting move starts from a chosen high potential energy configuration (shooting point $(\mathbf{x}_\mathrm{sp}, \mathbf{v}_\mathrm{sp})$) close to the TS  with randomly assigned velocities.
The system commits/relaxes towards one of the basins.
If a shooting move commits to basin A, all the configurations between the shooting point and the endpoint are labeled as A (n=0).
For configurations close to transition state hypersurfaces, the same configuration can appear in several trajectories with different $(\mathbf{x}_\mathrm{sp}, \mathbf{v}_\mathrm{sp})$ that commit to different basins and thus be labeled differently.
This method can then be used to statistically sample the basin label for each configuration.

The resulting arrival probability learned with Eq. \ref{eq:crossentropy} is different from the committor $q$.
Because the basin label of non-shooting point configurations depends on the shooting point $(\mathbf{x}_\mathrm{sp}, \mathbf{v}_\mathrm{sp})$,
the learned probability distribution $p_2$ is
\begin{equation}
    p_2(n=1|\mathbf{x})= c\int_\Omega q(\mathbf{x}_\mathrm{sp}, \mathbf{v}_\mathrm{sp}) d\mathbf{x}_\mathrm{sp} d\mathbf{v}_\mathrm{sp}
    \label{eq:tildep}
\end{equation}
where $\Omega$ contains the starting configurations that lead to a trajectory that arrives at $\mathbf{x}$ before committing to a basin
and $c$ is the normalizing factor.
However, in case of a one-dimensional reaction tube, it can be shown that $p_2$ is a monotonic transformation of $q$ (in Supplementary Section E).
More importantly, this monotonic dependency can be passed to the latent space variable $\xi$ when the classifier monotonically transform $\xi$ to $p_2$ with $g(\xi)=p_2$,
In other words, $\xi$ can be monotonic transformation of $q$ if $g$ is a monotonic function.
Therefore, all the classifiers used in section III and IV will be purely linear.

Nonetheless, $p_2$ is still a good approximation to $q$ numerically in the TS region where $q\approx 0.5$, as confirmed by the results in section III and IV.
This is because the majority of the data around the TS region are shooting point configurations.
Thanks to this correlation, the decision boundary of our classifier $p_2=0.5$ is close to actual transition states, where $p_1=0.5$.
In the remaining region where $q$ is close to 0 (or 1), $p_2$ suffers from the same numerical accuracy problem as $q=p_1$. 
Because both $p_1$ and $p_2$ will be either 0 or 1 in these regions,
the learned CVs may not be able to differentiate different intermediate states from the basin to the TS, mapping them to the same CV value.
But later in the discussion, it will be shown that the potential energy label can help remedy such numerical issues by separating these intermediate states while preserving their order in the reaction progress.

\subsection{Implementation}
The neural network and training framework are implemented with Tensorflow 1.14\cite{tensorflow2015-whitepaper}.
The encoder, classifier and PEP are then trained together
by the Adam algorithm\cite{2014arXiv1412.6980K} with a slowly decreasing learning rate
which is reduced by 5\% every 20 epochs.
The number of nodes used for the encoder, PEP are listed in 
Supplementary Table I.
In the first 100 epochs, the prefactors $c_p$ are varied randomly with a uniform distribution as followed,
\begin{equation}
c_\mathrm{part} \sim \mathrm{Unif}(0, M_\mathrm{part}), \quad \quad \mathrm{part} = \mathrm{clf}, \mathrm{pe}, \mathrm{reg}.
\end{equation}
The magnitude $M_\mathrm{part}$ is chosen such that 
$L_\mathrm{clf}$ and $L_\mathrm{pe}$ contribute equally to the initial loss function value, 
while regulation terms contribute less than 5\%.
The choice of starting learning rates, numbers of epochs and the magnitudes $M_\mathrm{part}$ 
are listed in Supplementary Table II.

More details of the implementation and the training protocol can be found in Supplementary Section A and the Harvard Dataverse repository\cite{nncvrepo}. 

\section{Case studies on model systems}

In this section, the multitask learning framework is applied to three model systems: a 5D M{\"u}ller-Brown model, a 5D three-well model and alanine dipeptide in vacuum.
These three model systems all have well-defined ideal CVs that can be used to accelerate sampling and compute accurate free energies.
Our goal is to examine the model's ability to learn complex reaction paths, and hence the CVs in these models are non-linearly related to the input features, i.e. Cartesian coordinates.

\subsection{5D M{\"u}ller-Brown and three-well model}

We first consider two variations of a model parameterized by five-dimensions.
The potential energy $V_\mathrm{5d}(x_1, x_2, \dots, x_5)$ is taken as a nonlinear transformation from a two-dimensional function $V_\mathrm{2d}$ as follows,
\begin{align}
V(x_1, x_2, \dots, x_5) & =  V_\mathrm{2d}(\tilde{x}, \tilde{y}) \\
\tilde{x} & = \sqrt{x_1^2+x_2^2+10^{-7}x_5^2} \label{eq:tildex}\\
\tilde{y} & = \sqrt{x_3^2+x_4^2}.
\label{eq:tildey}
\end{align}
The 2D subspace $(\tilde{x}, \tilde{y})$ unambiguously determined the potential energy, while the derived 5D model is made largely degenerate in energy.

\begin{figure}[h]
\caption{The potential energy landscape of (a) M{\"u}ller-Brown model and (b) three-well model in the $(\tilde{x}, \tilde{y})$ subspace
\label{fig:energy_landscape}}
\includegraphics[width=0.8\linewidth]{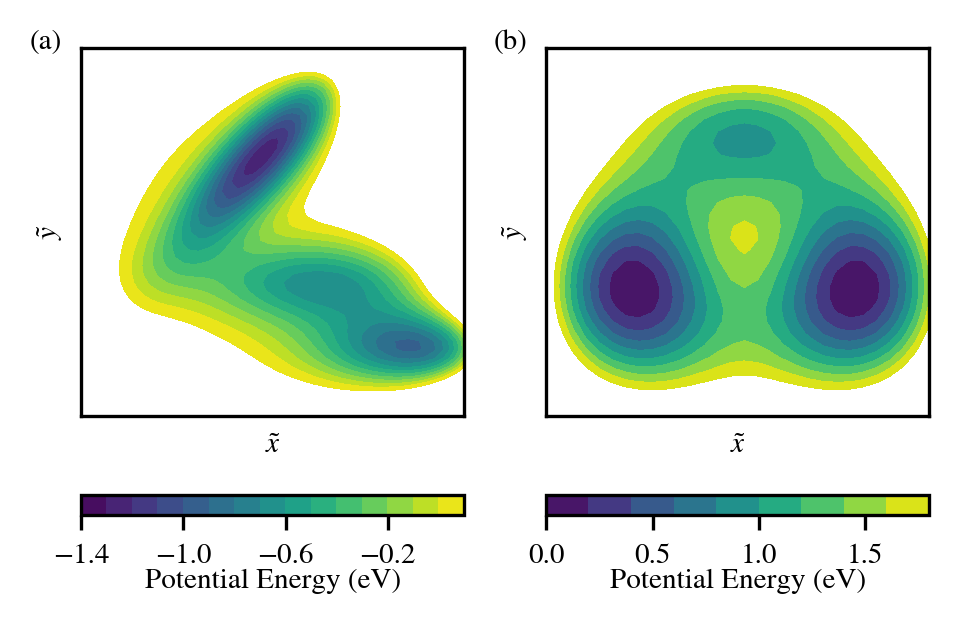}
\end{figure}

The underlying 2D potential function is defined as,
\begin{align}
V_\mathrm{2d}(\tilde{x}, \tilde{y}) & = \sum_{i=1}^4 A_i \exp\biggl[\alpha_i(\tilde{x}-a_{i})^2 + \beta_i(\tilde{x}-a_{i})(\tilde{y}-b_{i}) \notag \\
&\qquad + \gamma_i(\tilde{y}-b_i)^2) \biggr] - D(\tilde{x}-d)^3 \notag \\
&\qquad - E(\tilde{y}-e)^3.
\label{eq:mueller_brown}
\end{align}
Two sets of coefficients, listed in Supplementary Table III and IV,  are used to generate a double-well model (so-called M{\"u}ller-Brown model) and a three-well model.
These coefficients are originally devised by M{\"u}ller and Brown\cite{muller_location_1979} and Metzner et al.\cite{Metzner_Schutte_Vanden-Eijnden_2006},
but we scaled them up to increase the reaction barrier height for dynamics at the temperature $T=300$ K.
The M{\"u}ller-Brown model (Fig.\ref{fig:energy_landscape}(a)) has two metastable states (A and B) and one minimum energy path between the two basins with a transition barrier of 0.9 eV (~30$k_BT$) from A to B and 0.6 eV (~23$k_BT$) from B to A.
In the three-well model (Fig.\ref{fig:energy_landscape}(b)),
two basins (A and B) have lower potential energy than the third basin, C.
The transition barrier is 1.0 eV from A/B to C and 1.2 eV from A to B.

Even though the basins for these two models are separable in the 2D $(\tilde{x}, \tilde{y})$ subspace, their structure in the 5D space is obscured by the non-linear embedding and degeneracy.
For example, a linear path in the $(\tilde{x}, \tilde{y})$ subspace (Fig.\ref{fig:energy_landscape}(e)) is equivalent to a hypersurface in the 5D space (Fig.\ref{fig:energy_landscape}(c-d)).

For each model, 300,000 configurations were collected from MD trajectories near the basin (without transitions) and trajectories from transition path sampling, shown in Fig. \ref{fig:5d_MB_train_learned}(a) and (b), respectively.
The simulation details are documented in Supplementary Section C.
In order to test how well the model is able to learn and generalize the transition dynamics, the training and test sets are initialized with different $x_1/\tilde{x}$, $x_3/\tilde{y}$ ratio, such that the two sets have no overlap in the 5D space but overlap significantly in the $(\tilde{x}, \tilde{y})$ subspace (Fig. \ref{fig:5d_MB_train_learned}(c)).
The intention is to test whether the algorithm correctly reduces dimensionality to discover the ``true'' CV which is given by the 1D minimum energy path connecting the basins in the first model and the 2D $(\tilde{x}, \tilde{y})$ subspace in the second model.
The expectation is that if the low-dimensional manifold is identified correctly from the training set, the model will be able to generalize (achieve low prediction error) on the test set even in the presence of degeneracy.

\begin{figure}[h]
\caption{5D M{\"u}ller-Brown model.
(a) Training data plotted in $(\tilde{x}, \tilde{y})$ subspace. The configurations labeled with Basin A and B are colored as red and blue, respectively. 
The dots with paler colors are obtained from TPS simulations while the darker ones are from simulations trapped at the basins.
The background color contours depict the potential energy $V_\mathrm{2d}(\tilde{x}, \tilde{y})$. 
(b) Training and test sets in $(x_1, x_2)$ subspace, colored grey and orange, respectively.
(c) Contour of $\xi$ in the $(\tilde{x}, \tilde{y})$ subspace.
The $\xi$ value of each $(\tilde{x}, \tilde{y})$ value is averaged from five sets of $(x_1, x_2, ..., x_5)$.
The black dots are configurations from the test set.
The grey lines are true potential energy contours.
(d) The predicted/actual potential energy ($\tilde{V}$/$V$) and basin label ($\tilde{n}$/$n$) as a function of $\xi$.
\label{fig:5d_MB_train_learned}}
\includegraphics[width=\linewidth]{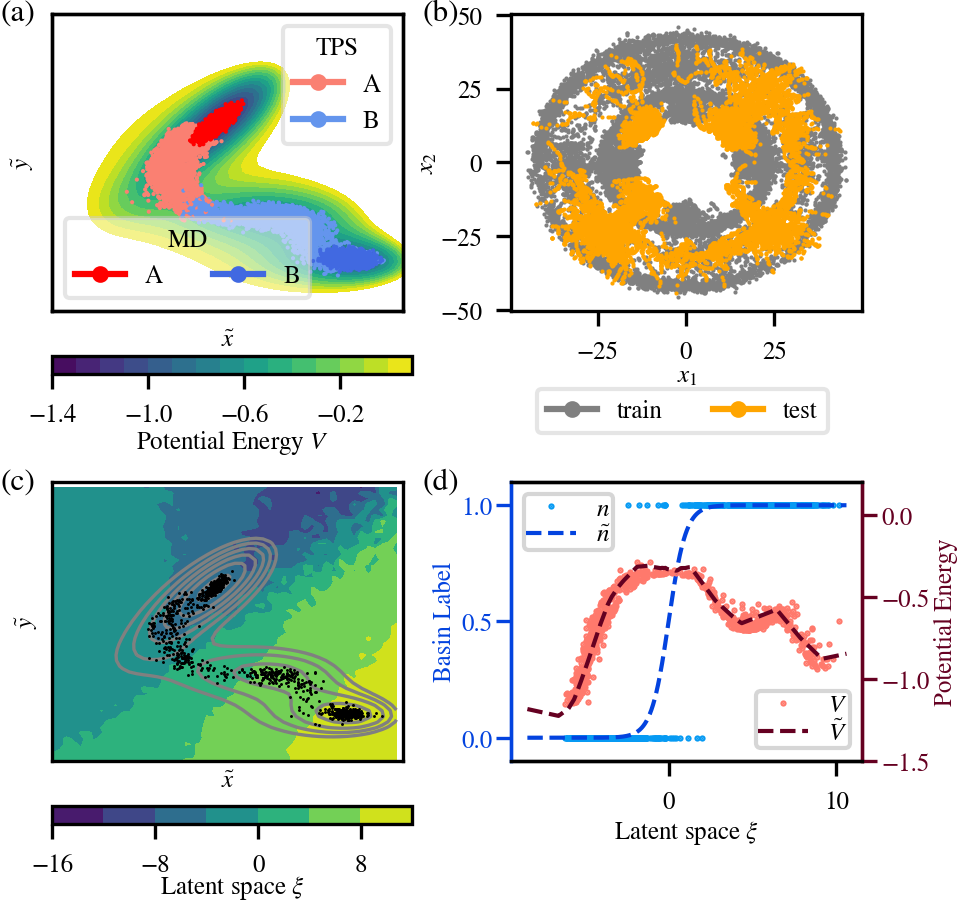}
\end{figure}

For the M{\"u}ller-Brown model, the latent space variable $\xi$ is chosen to be one dimensional.
After training, the classifier achieves an accuracy of 98\%  on the test set.
As shown in Fig. \ref{fig:5d_MB_train_learned} (d),
the 2\% mis-classified configurations are located around the transition state $\xi=0$.
This misclassification is hard to avoid,
due to the slow variation of the potential energy landscape around the transition ridge. 
A slight velocity change can lead the system to a different basin, and thus the configurations around that region can be labeled as both A and B.
Fig. \ref{fig:5d_MB_train_learned}(b) shows that there is no clear separation of label A configurations and label B configurations even in the $(\tilde{x}, \tilde{y})$ subspace.
Thus, this vague separation boundary is kept in the latent space $\xi$.
The PEP predicted potential energy $\tilde{V}$  closely follows the ground-truth values with a test set mean absolute error (MAE) of 0.04 eV  as shown on Fig. \ref{fig:5d_MB_train_learned})(b).
The change of $\tilde{V}$ along $\xi$ is similar to the actual potential energy $V$ change along the path from A to B. In particular, $V$ is maximized at the decision boundary of the classifier ($\xi=0$ in \ref{fig:5d_MB_train_learned})(b)).

As shown in Fig. \ref{fig:5d_MB_train_learned} (a), $\xi$ is relatively smooth and, more importantly, $\xi$ is seen to monotonically tracks to the reaction progress.
Especially in the area covered by the test set, the $\xi$ contours are perpendicular to the reaction path and tangential to the potential energy isosurface.

\begin{table}[]
\centering
\begin{tabular}{ccc}
\hline
case & $\Delta F_\mathrm{A\rightarrow B}$ & $F_\mathrm{B}-F_\mathrm{A}$ \\
\hline
$(\tilde{x}, \tilde{y})$  & $0.89\pm 0.02$ & $0.51\pm 0.02$ \\
$\xi$ & 0.94 & 0.43 \\
\hline\\
\end{tabular}
\caption{Reaction Free energy from basin A to basin B of the 5D M{\"u}ller-Brown model 
computed with umbrella sampling along the ``true'' collective variables 
$\tilde{x}, \tilde{y}$ and the latent space variable $\xi$.
Units: eV.
\label{tab:5dMB_free_energy}}
\end{table}

Umbrella sampling is employed to compute the free energy profile along $\xi$ and in $(\tilde{x}, \tilde{y})$ subspace,
which is summarized in Table \ref{tab:5dMB_free_energy}, with detailed plots given in Supplementary Fig. 1.
Free energies computed with the latent space variable $\xi$ are reasonably close to the one estimated with the ``true'' CVs with an error of 0.05-0.08 eV.

\begin{figure}[h]
\caption{5D three-well model.
(a) Training data in $(\tilde{x}, \tilde{y})$ subspace.
The configurations labeled with Basin A, B and C are colored as blue, orange and green, respectively.
The dots with paler colors are obtained from TPS simulations while the darker ones are from MD simulations.
The background color contours depict the potential energy $V_{2d}(\tilde{x}, \tilde{y})$.
(b) Training and test sets in $(x_3,x_4)$ subspace, colored grey and orange, respectively.
(c-d) Potential energy and basin labels in the latent space $(\xi_1, \xi_2)$.
The background colors represent (c) the predicted potential energy $\tilde{V}$ and (d) the predicted basin $\tilde{n}$.
The scatter points represent the test set, colored by (c) actual potential energies $V$ and (d) actual basin labels $n$.
\label{fig:5d_3h_train_learned}}
\includegraphics{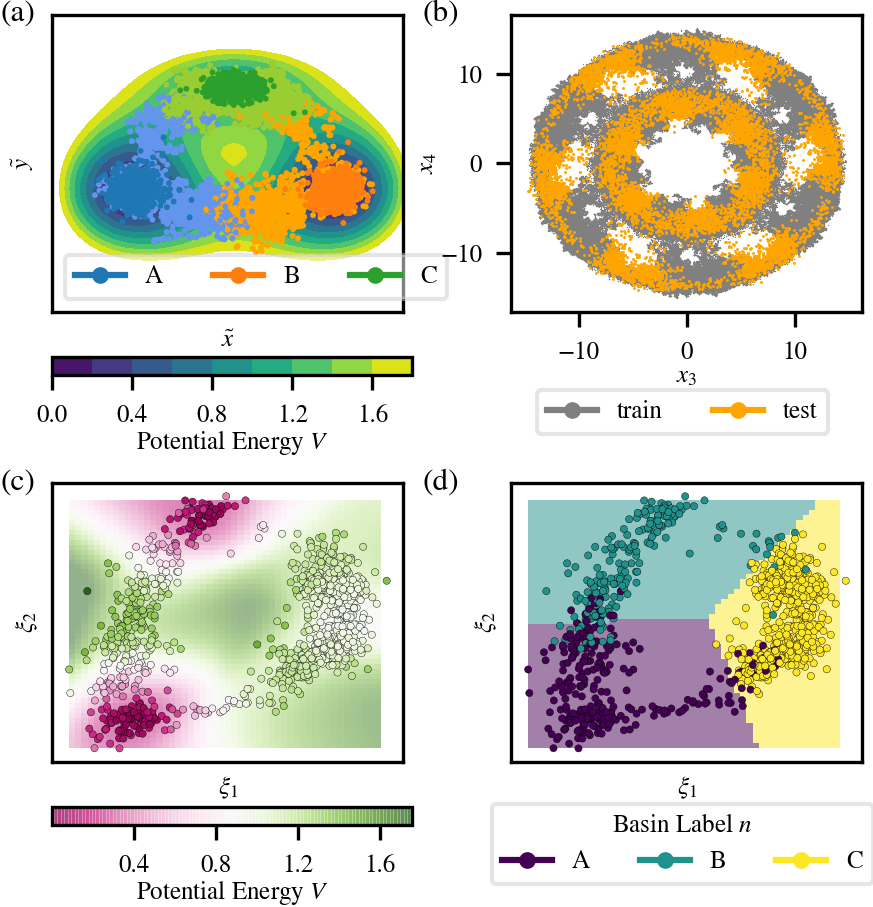}
\end{figure}

For the three-well model, a 2-D latent space $(\xi_1, \xi_2)$ is learned because a single dimension is not enough to differentiate three different transition paths ( A $\leftrightarrow$ B, A $\leftrightarrow$ C, B $\leftrightarrow$ C).
The trained classifier is able to divide the $(\xi_1, \xi_2)$ subspace into three regions with an accuracy of 90\%, 
as shown in Fig. \ref{fig:5d_3h_train_learned} (d).
In this case, the boundaries between these regions are linear because a linear classifier is used.
It is worth noting that this 90\% accuracy is close to its theoretical limit because the potential energy landscape is relatively flat around basin C.
This is because when using the ``true'' CVs $(\tilde{x}, \tilde{y})$ as the input, the classification accuracy is seen to be limited to 90\%. The PEP prediction for the three-well model has a mean absolute error of 0.07 eV.
As shown in Fig. \ref{fig:5d_3h_train_learned}(c), the PEP reflects the energy rise and fall along all three transition paths.

\subsection{Alanine dipeptide}

The multitask learning algorithm is also applied to a real molecular system, alanine dipeptide in the vacuum.
This 22-atom molecule is often used as a model system to demonstrate protein folding and to test dimension reduction algorithms\cite{branduardi_b_2007}.
Compared to the above toy models, alanine dipeptide is more complex 
due to its higher input dimension (66 Cartesian coordinates).
To train the multitask model, the configurations are shifted and rotated such that the center carbon atom locates at (0, 0, 0), the two connecting C atoms lie on the $x-y$ plane, and one of them lies on the $x$-axis. 
Thus, the input feature dimension is, in fact, 63.

This molecule has three metastable states in vacuum, C7$^\mathrm{ax}$, C7$_{eq}$ and $\beta$ states\cite{Strodel_Wales_2008}.
The bottom of these three metastable states can be identified in the 2D Ramachandran plot using two backbone dihedral angles $\phi$ and $\theta$ as coordinates, well-known widely used set of good CVs for this system.
At temperatures below 150 K, no transition between states occurs in a one-nanosecond long unbiased MD simulation (details in Supplementary Fig. 2.
 C7$_{eq} \leftrightarrow \beta$ transitions are observed above 200 K.
As the temperature increases, the C7$_{eq}$-$\beta$ basin and the C$_7^\mathrm{ax}$ basin grow larger.
Above 600 K, some transition events from and to the C$_7^\mathrm{ax}$ can occur within 1 ns.

\begin{figure}[h]
\caption{Illustration of training data of alanine dipeptide in the torsion angles ($\phi, \theta$) subspace. The molecule structure of alanine dipeptide is plotted as a subset in (a). $\phi$ and $\theta$ are two torsion angles of the C-N chain. Each point in the plot represents one atomic configuration and they are colored by (a) the potential energy and (b) the basin label.\label{Ala_dip_demo}}
\includegraphics{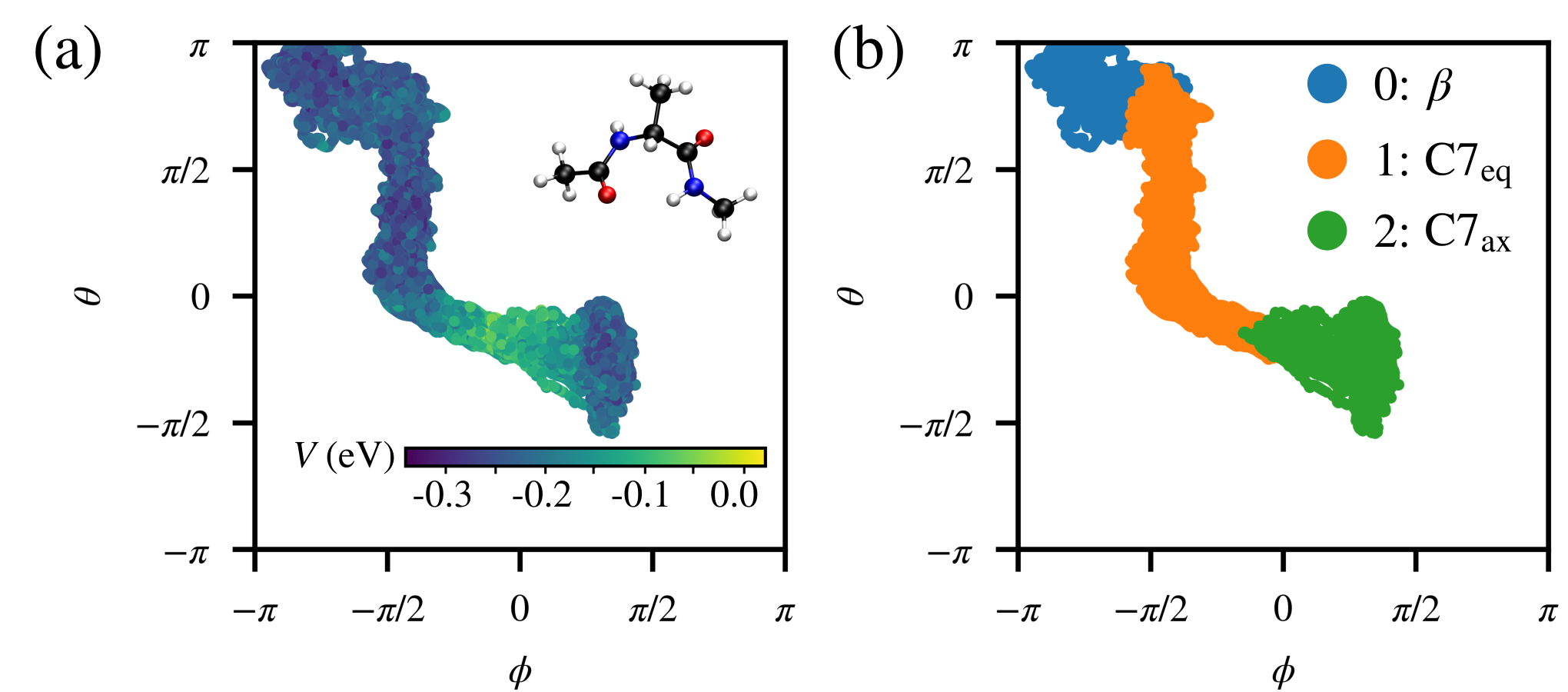}
\end{figure}

\begin{table}[]
\centering
\begin{tabular}{ccc}
\hline
case & $\Delta F_{C7_{eq}\rightarrow C7_{ax}}$ & $F_{C7_{eq}}-F_{C7_{ax}}$ \\
\hline
$\phi$  & 0.32 & 0.03 \\
$\xi$ & 0.41 & -0.001\\
\hline\\
\end{tabular}
\caption{Reaction Free energy from C7$_{eq}$ toC7$_{ax}$ of alanine dipeptie 
computed with umbrella sampling along the ``true'' collective variables 
$\phi$ and the latent space variable $\xi$ at 50 K.
Units: eV.
\label{tab:ala_dip_free_energy}}
\end{table}

Using the 700 K transition events as seeds, transition path sampling can find two different transition paths connecting these three metastable states at 120 K.
The transition path ensemble is visualized in Ramachandran plot in Fig. \ref{Ala_dip_demo}.
The C7$_{eq} \leftrightarrow \beta$ transition has a lower potential energy at the saddle point than that of the C7$_{eq} \leftrightarrow \mathrm{C7}_{ax}$ transition.
The test set is obtained in a separate TPS simulation at a slightly lower temperature of 100 K.

\begin{figure}[h]
\caption{ (a) Torsion angles $\phi$ and $\theta$, (b) predicted basin label $\tilde{n}$, actual basin label $n$, predicted potential energy $\tilde{V}$ and actual potential energy $V$ as a function of latent space variable $\xi$. The training data is obtained at 120 K and the test data is obtained at 100 K. \label{fig:ala_dip_1d_line}}
\includegraphics[width=\linewidth]{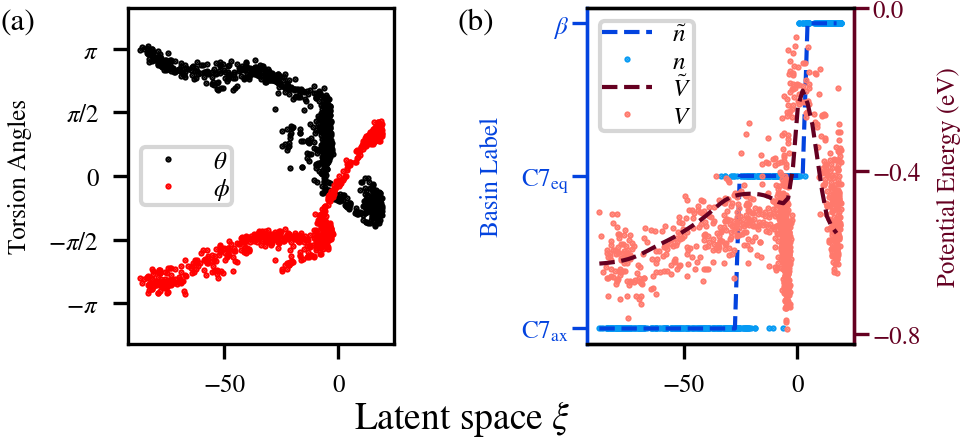}
\end{figure}

The multitask network learned a one-dimensional CV, $\xi$, using the atomic Cartesian coordinates as input.
The classifier achieves 86\% accuracy and the PEP network predicts potential energy with a  mean absolute error of 0.1 eV.
Similar to the 5D models, $\xi$ can separate the three basins and reflect the change of potential energy from the saddle points to the basin, as shown in Fig. \ref{fig:ala_dip_1d_line}.
Moreover, when trained using the multitask architecture, the encoder can learn the important structural features, the $\phi$ and $\theta$ dihedral torsion angles, from the Cartesian coordinates.
This is reflected in Fig. \ref{fig:ala_dip_1d_line}(a) which shows that the leaned $\xi$ is smoothly connected to the two torsion angles $\phi$ and $\theta$. 
In contrast, we find that a variety of single-task learning procedures results in much less smooth connection between $\xi$ and the $(\phi, \theta)$ set (see Section IV and Supplementary Section IV.2).

Compared to the previous 5D models, the classification and potential energy accuracy is lower for alanine dipeptide due to a larger ratio between thermal fluctuations and the reaction barrier.
As a result, many configurations around the $\mathrm{C7}_\mathrm{eq}$ basin are mapped to a small range of $\xi$ (Fig. \ref{fig:ala_dip_1d_line}(a)).
However, the $\xi-\phi$ and $\xi-\theta$ correlation is still relatively smooth in this region.
We also note that the training can be much easier with training data obtained using a lower temperature of 50 K 
since thermal fluctuations are smaller and potential energy describes better with the reaction progress.

\begin{figure}[h]
\caption{ Umbrella sampling at 50 K for C7$_\mathrm{eq}$-C7$_\mathrm{ax}$ transition.
The model is trained with 50 K data.
(a) Sampled atomic configurations plotted in the torsion angles $(\phi, \theta)$ subspace;
colored by latent space variable $\xi$.
(b) free energy and potential energy profile along latent space variable $\xi$. The blue line represents the free energy and the orange scatter points represent the potential energies.}

\label{fig:ala_dip_umb}
\includegraphics[width=\linewidth]{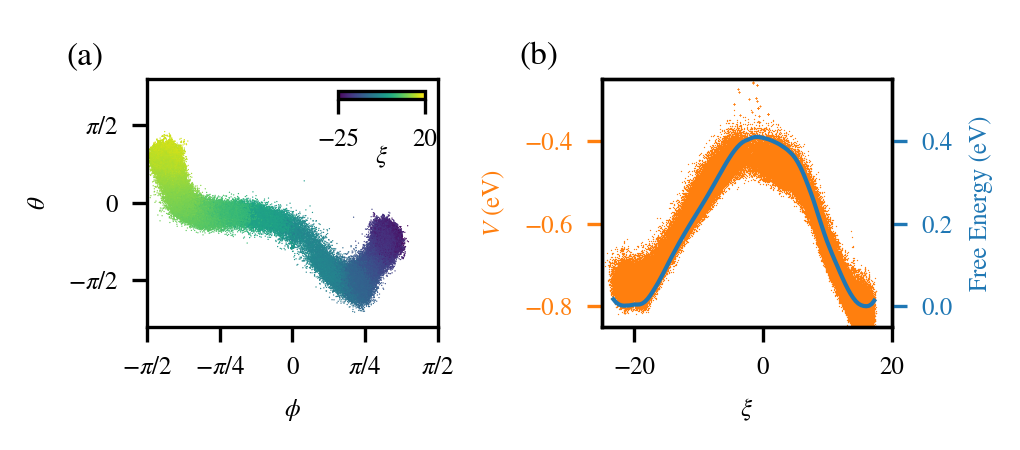}
\end{figure}

The learned latent space variable $\xi$ from 50 K training data is then used as a reaction coordinate for the reaction between C7$_\mathrm{eq}$ and C7$_\mathrm{ax}$.
In this reaction, torsion angle $\phi$ is enough to describe the reaction progress,
and thus only $\phi$ is used as the ``true'' CV reference.
Umbrella sampling is employed to sample the free energy landscape using LAMMPS\cite{Plimpton_1995} and PLUMED\cite{Bonomi_2009} codes with an additional interface to load the Tensorflow neural network \cite{TF_PLUMED}.
The reweighing and free energy profile are computed with the bin-less multi-state free energy estimation method \cite{Tan_Gallicchio_Lapelosa_Levy_2012} and UWHAM \cite{Zhang_Arasteh_Levy_2019}.
As shown in Fig. \ref{fig:ala_dip_umb} (a), the umbrella sampling trajectories sample the bottom of the basins and reaction path C7$_{eq} \leftrightarrow \mathrm{C7}_{ax}$.
And the free energy profile integrating along $\xi$ yields an accurate reaction free energy profile,
compared to the reference 
``true''
collective variable $\phi$ (Table \ref{tab:ala_dip_free_energy}).
In Fig. \ref{fig:ala_dip_umb}(b), the potential energies around the $\xi=0$ coordinate spikes more than other regions because the umbrella sampling constraints push the system to explore more high energy states.

\section{Discussion}

The key to efficient collective variables learning
is to combine the dimensionality reduction by the encoder with the downstream parts handling tasks T2 and T3 discussed in Section II.
To compare the performance of all three models discussed in section III, two single-task neural networks are trained: with only the classifier or the PEP downstream parts. 
For brevity sake, only the case of the 5D M{\"u}ller-Brown model is presented in the main text,
and the remaining data sets are left to Supplementary Sections C and D (Supplementary Fig. 3-5.

\begin{figure*}[ht]
\caption{ \label{5d_1d_contour} Comparison among different single-task architectures consisting of: (a,e) an encoder and a classifier, trained with $L_\mathrm{clf}$; (b, f) an encoder and a PEP, trained with $L_\mathrm{pe}$; (c, g) an encoder, trained with $L_\mathrm{pe}^\mathrm{pair}$; (d,h) an encoder and a decoder, trained with $L_\mathrm{reconst}$. (a-d) The spatial distribution of latent space variable $\xi$ in the $(\tilde{x}, \tilde{y})$ subspace. The black dots represent the location of training and test set and the background contours are colored by the $\xi$ value. We note that a single $(\tilde{x}, \tilde{y})$ value can correspond to many $(x_1, \dots, x_5)$ values. For each $(\tilde{x}, \tilde{y})$ in these plots, only one set of $(x_1, \dots, x_5)$ is randomly chosen to satisfy Eq. \ref{eq:tildex} and \ref{eq:tildey}. (e-h) predicted potential energy $\tilde{V}$, predicted label $\tilde{n}$ as a function of $\xi$.}
\includegraphics{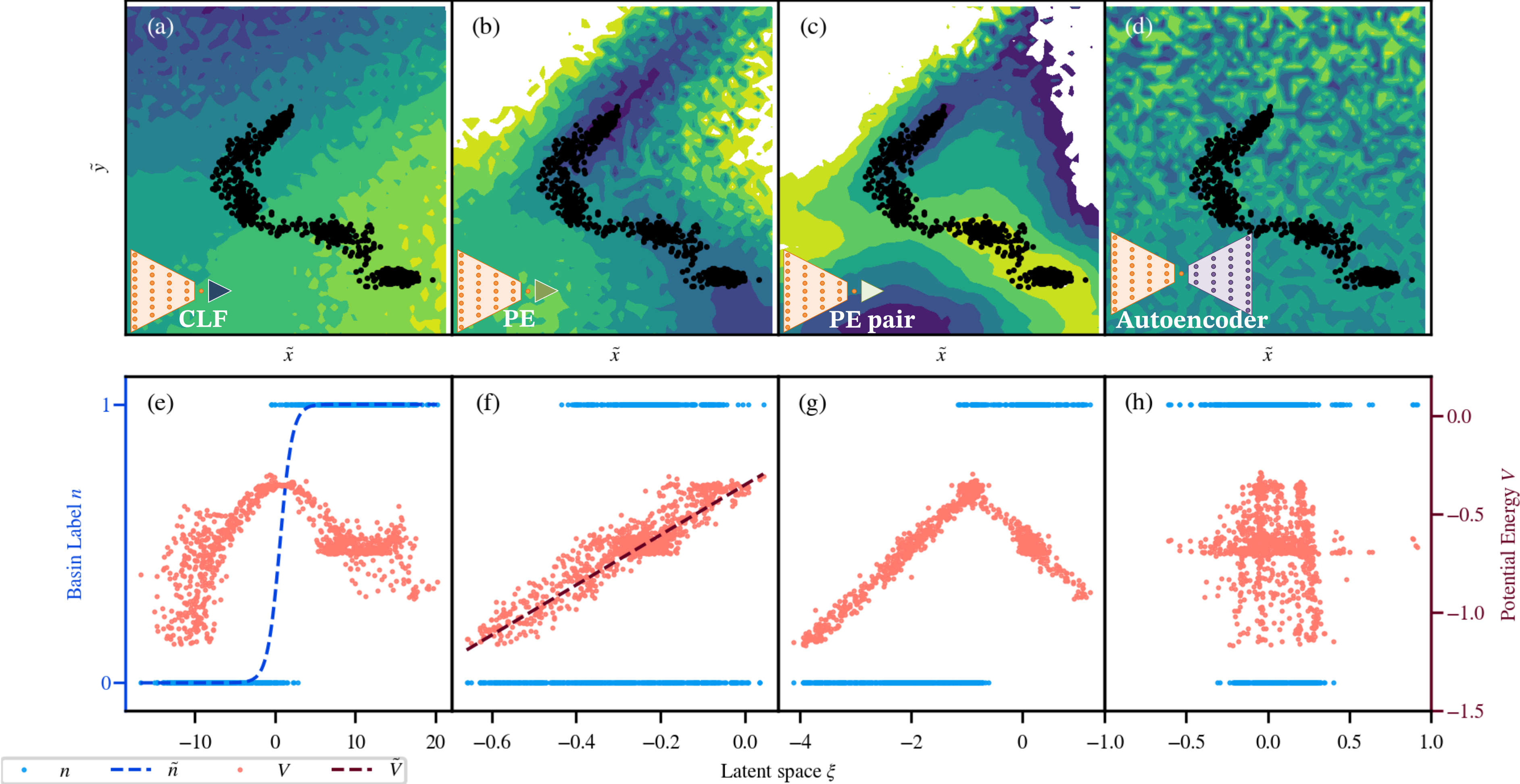}
\end{figure*}

In the first single task learning setup, the neural network has an encoder of the latent space $\xi_\mathrm{clf}$ and a classifier trained with only the $L_\mathrm{clf}$ loss function. 
The latent space $\xi_\mathrm{clf}$ in this case can still identify and separate the two basins 
with 95\% accuracy, as shown in Fig. \ref{5d_1d_contour}(a, e).
Around the transition states, the potential energy $V$ is sharply concentrated around its conditional mean on $\xi_\mathrm{clf}$,
and its contour is perpendicular to the reaction path 
around the saddle point region.
However, outside of the transition region,
the contour is tangential to the reaction path and different energy states mixed at the same $\xi_\mathrm{clf}$ value.
The free energy derived from this $\xi_\mathrm{clf}$ is 0.52 eV, 
prominently lower than the ground truth value 0.89 eV.
The free energy difference between basin A and B is also greatly under-estimated.

The mixing of high energy and low energy states and underestimated reaction free energy 
indicate that 
the single task network fails to learn the reaction path in underlying $(\tilde{x}, \tilde{y})$ subspace,
and $\xi_\mathrm{clf}$ cannot preserve the reaction progress.
It misses the nuance of the reaction progress exactly because of the numerical accuracy issue affecting estimation of the committor mentioned in section II.C, i.e. the classifier function has very small variation close to the basins.
This problem is less severe for the region around the transition state
because most of the training data around that area are generated from the TPS shooting point configurations.
That is why the potential energies around the transition state ( $\xi_\mathrm{clf}\approx0$) in Fig. \ref{5d_1d_contour}(e) are closely correlated with $\xi_\mathrm{clf}$.

Next, we consider a single-task learning framework where the network has an encoder and a PEP, trained with only $L_\mathrm{pe}$ loss.
The contours of $\xi$ in Fig. \ref{5d_1d_contour}(b) are close to the true potential energy.
However, Fig. \ref{5d_1d_contour}(f) shows that while the encoder clearly orders configurations by their potential energy in the latent space, it assigned many the configurations from two
different basins onto the same values of $\xi$.

The success of multitask learning originates from the synergistic effect among all parts of the loss function.
Effect of the $L_\mathrm{clf}$ loss dominates around
the transition state hypersurfaces ($\xi\approx 0$) and
removes the energy degeneracy across basins by separating them in the latent space.
Simultaneously, minimizing the $L_\mathrm{pe}$ loss tends to order the configurations by potential energy so that the reaction progress from the bottom of the basin towards the transition states is preserved.


There is the freedom to choose the exact expressions for $L_\mathrm{pe}$ and $L_\mathrm{clf}$
as long as they accomplish tasks T2 and T3 listed in Section IIA.
In fact, T3 (preserving atomic structural evolution) can be achieved by any loss function and architecture that captures information about the proximity between configuration along the reaction path.
We demonstrate below a successful example with an alternative form of the loss depending on the potential energy $L_\mathrm{pe}^{\mathrm{pair}}$ and an unsuccessful example of an autoencoder reconstruction loss $L_\mathrm{reconst}$.

In the first example, potential energy is used to measure the proximity of configurations, whereby the pairwise distance between two configurations $i$ and $j$ in the latent space 
is trained to match their potential energy difference.
Thus, $L_\mathrm{pe}^{\mathrm{pair}}$ is taken as the L2 norm 
for differences between $V_i-V_j$ and $d^{(l)}_{ij}$, written as
\begin{equation}
\label{eq:pe_loss2} 
L_\mathrm{pe}^{\mathrm{pair}} = \sum_{\left\{i, j | n_i=n_j \right\}} u_{ij}
[(d^{(l)}_{ij})^2-(V_i - V_j)^2  ]^2  
\end{equation}
Here $V_i-V_j$, is the potential energy difference
between two data points $i$ and $j$ and $d^{(l)}_{ij} = \|\xi_i-\xi_j\|$ is the pairwise Euclidean 
distance in the reduced $l$-dimensional latent space.
Only $(i,j)$ pairs where both points are 
within the same basin class are considered.
Unlike $w_i$ in Eq. \ref{eq:pe_loss1}, 
which is defined for a single data point, 
the weight $u_{ij}$ is defined for a pair of data points,
\begin{equation}
u_{ij} = 1 - s\left(\frac{\left|V_i - V_j\right|-V_0}{\Delta V}\right) 
\label{eq:pe_weight2}
\end{equation}
where $s$ is a sigmoid function. 
This weight drops to zero 
when the potential energy difference of the pair 
is much greater than $V_0$ ($\left|V_i - V_j\right| \gg V_0$).
The parameter $\Delta V$ is used to control how fast the weight drops 
to zero around $V_0$.

In the second example, an autoencoder \cite{baldi2012autoencoders} is tested, which maps atomic configuration to the latent space variable $\xi_\mathrm{reconst}$ with an encoder and then maps $\xi_\mathrm{reconst}$ onto 
a reconstructed configuration $\tilde{\mathbf{x}}_i$ with a decoder. 
Autoencoders are often used for dimension reduction and manifold learning.
It is generally believed that the latent space can preserve the proximity of configurations by minimizing the Euclidean distances $d_{ij}^{(3N_a)}$ between the original and reconstructed atomic configurations.
The reconstruction loss$L_\mathrm{reconst}$ is defined as follow,
\begin{equation}
L_\mathrm{reconst}
= \sum_i \left | \mathbf{x}_i-\tilde{\mathbf{x}}_i\right |^2 
\end{equation}
where $N_a$ is the number of atoms in the atomic configurations. 

For $L=L_\mathrm{pe}^{\mathrm{pair}}$, 
Fig. \ref{5d_1d_contour}(g) shows that the encoder still orders the configurations by potential energy in the latent space, 
with the additional feature that the two basins are separated,
thanks to the fact that only same-basin pairs are used in Eq. \ref{eq:pe_loss2}.
The resulting $\xi_\mathrm{pe}^\mathrm{pair}$ is very closely correlated to the actual potential energy, as shown in $(\tilde{x}, \tilde{y})$ in Fig. \ref{5d_1d_contour}(c).
More interestingly, such correlation extends beyond the training and test set region in the $(\tilde{x}, \tilde{y})$ space.
In this sense, it is better than the multitask latent space because of its robustness.

However, this observed robustness is purely fortuitous.
For the 5D three-well model, $L_\mathrm{pe}^{\mathrm{pair}}$ mixes basins A and B as seen in Supplementary Fig. 3(g).
Also, Supplementary Fig. 4-5(g) show that for alanine dipeptide, $L_\mathrm{pe}^{\mathrm{pair}}$ mixes the class labels for all three basins.
But this problem can be remedied by introducing additional terms in the loss function to separate pairs of configurations belonging to different basins.
Eq. \ref{5d_1d_contour} is just an example. 

For the autoencoder with $L_\mathrm{reconst}$, the basin labels and potential energies are entangled in the latent space $\xi_\mathrm{reconst}$.
From Fig. \ref{5d_1d_contour}(d) it appears that $\xi_\mathrm{reconst}$ has negligible correlation with the ``true'' CVs $\tilde{x}$ or $\tilde{y}$.

A successful loss function must guide the latent space to preserve the proximity between configurations in the reaction progress.
Even though the autoencoder directly uses Euclidean distances $\left | \mathbf{x}_i-\tilde{\mathbf{x}}_i\right |$,
it can be overwhelmed by the large-amplitude low energy perturbations that do not correlate with reaction progress.
In the two 5D models, points that are close in the 2D $(\tilde{x}, \tilde{y})$ subspace can be far away from each other in the 5D space.
In particular, $x_5$ has little impact on $\tilde{x}$ and the potential energy, but its fluctuation can dominate the 5D Euclidean distance between configurations.
Therefore, $L_\mathrm{reconst}$ loss is likely to fail due to the interference of $x_5$.
In fact, for the two 5D models, almost all standard dimension reduction techniques that are based solely on Euclidean distances in the 5D space are expected to struggle.
Using potential energy as the distance function can remove the influence of these large amplitude low-energy fluctuations.
It is especially useful for energy-activated transitions, because potential energy is closely related to the reaction progress in the configuration space around the reaction path. That is why including $L_\mathrm{pe}$ or $L_\mathrm{pe}^{\mathrm{pair}}$ in the multitask joint loss along with the basin label information tends to work well.

In addition, the limited dimensionality of the latent space may inhibit the performance of autoencoders.
Autoencoders may have a better chance if the dimensionality of $\mathbf{\xi}_\mathrm{reconst}$ exceeds the intrinsic dimension of the data manifold \cite{Dai_Wang_Aston_Hua_Wipf_2018}.
For example, Chen et al.\cite{chen_collective_2018} used an autoencoder with a 2D latent space for exploring the energy landscape of alanine dipeptide, while Wang et al.\cite{Wang_Gomez-Bombarelli_2019} used a 9-dimensional latent space to coarse grain the same molecule.

The multitask framework introduced in this work can take training configurations from different types of simulations, as long as the basin class and potential energy labels are available.
The training data do not need to follow the Boltzmann distribution, as required by methods for finding the committor function, or be Markovian, as required by the methods for finding slow eigenmodes.
At the same time, it can also be generalized to accommodate additional downstream parts with other manifold learning loss functions that utilize time correlations of configurations.

We note that our learning objective is limited to reactions that involve a substantial change in potential energy.
It assumes the data is distributed around reaction tubes whose potential energy correlates well with reaction progress or around the bottom of basins whose potential energy does not vary significantly.
For diffusion-dominated processes or reactions with entropy bottlenecks, 
potential energy is not a good distance metric, and our learning procedure may not have advantages over existing methods.

\section{Summary}

In summary, we propose to use a multitask training algorithm to learn collective variables from configurations labeled by the basin class and potentials energy. 
These can be obtained, for instance, from MD trajectories and transition path sampling trajectories.
The neural network architecture contains an upstream encoder that maps atomic configurations onto a low-dimensional latent space, and two other downstream networks that predict the basin labels and potential energy from the latent space value,
which is optimized for the classification of configurations among the basins and the prediction of the potential energy.

The algorithm is applied to study a 5D M{\"u}ller-Brown model, a 5D three-well model and alanine dipeptide.
We show that due to the synergy in the multiple learning objectives, the multitask model can perform nonlinear dimensionality reduction and identify collective variables that represent well the reaction progress between the basins.
Finally, we demonstrate that the learned collective variables can be used in enhanced sampling methods, such as umbrella sampling, to obtain accurate free energy barriers. This approach opens possibilities for automated discovery of low-dimensional coordinates for describing a variety of chemical reactions and computing their rates.

\begin{acknowledgments}

The authors thank the helpful discussion with Patrick Riley, Ekin Dogus Cubuk, and Kai Kohlhoff.
L. S., S.B. and W. C. are supported by the Integrated Mesoscale Architectures for Sustainable Catalysis (IMASC), an Energy Frontier Research Center funded by the US Department of Energy (DOE), Office of Science, Office of Basic Energy Sciences under Award No. DE-SC0012573;
S. B. J.V and B. K. acknowledge partial support from Bosch Research.
J.V. is partially supported by the National Science Foundation (NSF), Office of Advanced Cyberinfrastructure, Award No. 2003725.
Y.X. is supported by the US Department of Energy (DOE) Office of Basic Energy Sciences under Award No. DE-SC0020128.
The training data is generated with Cori at the National
Energy Research Scientific Computing Center (NERSC), a
DOE Office of Science User Facility supported under Contract
No. DE-AC02-05CH11231, through allocation m3275.
And the models are trained on the FASRC Cannon cluster supported by the FAS Division of Science Research Computing Group at Harvard University and Longhorn in the Frontera project supported by Texas Advanced Computing Center (TACC) of the University of Texas at Austin, through allocation DMR20013.

\end{acknowledgments}

\bibliography{main}

\end{document}